%% file: main.tex
\documentclass{article}

\usepackage{hyperref}
\input{packages}
\input{custom}

\usepackage[accepted]{mlsys2024}

\mlsystitlerunning{Hierarchical Autoscaling for Large Language Model Serving}

\begin{document}

\twocolumn[
\mlsystitle{Hierarchical Autoscaling for \\ Large Language Model Serving with \sysname}
\mlsyssetsymbol{equal}{*}

\begin{mlsysauthorlist}
\mlsysauthor{Archit Patke}{uiuc}
\mlsysauthor{Dhemath Reddy}{uiuc}
\mlsysauthor{Saurabh Jha}{ibm}\\
\mlsysauthor{Chandra Narayanaswami}{ibm}
\mlsysauthor{Zbigniew Kalbarczyk}{uiuc}
\mlsysauthor{Ravishankar Iyer}{uiuc}

\end{mlsysauthorlist}

\mlsysaffiliation{uiuc}{University of Illinois at Urbana Champaign}
\mlsysaffiliation{ibm}{IBM Research}

\mlsyscorrespondingauthor{Archit Patke}{apatke@illinois.edu}



\mlsyskeywords{Machine Learning, MLSys}

\vskip 0.3in

\input{abstract}
]
\printAffiliationsAndNotice{}

\input{0000-intro_v2} 
\input{0001-background}

\input{0003-rpms}
\input{0003-local}
\input{0003-global}
\input{0006-results}
\input{0008-related}
\input{0009-conclusion}

\bibliographystyle{plainnat}
\bibliography{bibliography.bib}

\input{appendix}

\end{document}

%% file: packages.tex
\usepackage{textcomp}
\usepackage{pifont}
\usepackage{xspace}

\usepackage{makecell}
\usepackage{nicefrac}
\usepackage[utf8]{inputenc}
\usepackage[english]{babel}
\PassOptionsToPackage{bookmarks=false}{hyperref}
\usepackage{multirow, multicol, booktabs, tabulary, tabu, longtable, array, varwidth}
\usepackage{placeins, lipsum}
\usepackage[flushleft]{threeparttable}
\usepackage{tabularx}
\usepackage{amsmath}
\hypersetup{breaklinks=true}
\usepackage[labelfont=bf]{caption}

\usepackage{textcomp}
\usepackage[shortcuts,acronym]{glossaries}
\usepackage{soul, footnote, xargs}

\usepackage{amsmath}
\usepackage[utf8]{inputenc}

\usepackage{enumerate}

\usepackage[inline]{enumitem}
\setlist{noitemsep,nolistsep,leftmargin=*}

\usepackage{amsmath, amsfonts, amsthm, nicefrac}
\theoremstyle{definition}
\newtheorem{definition}{Definition}[section]

\usepackage{listings}

\usepackage[]{cleveref} 
\crefname{appsec}{Appendix}{Appendices}
\crefformat{definition}{Def. #2#1#3}
\crefformat{section}{Section #2#1#3}
\crefformat{subsection}{Section #2#1#3}
\crefformat{subsubsection}{Section #2#1#3}
\crefformat{equation}{(#2#1#3)}
\crefrangeformat{equation}{(#3#1#4--#5#2#6)}
\crefmultiformat{equation}{(#2#1#3)}{ and~(#2#1#3)}{, (#2#1#3)}{ and~(#2#1#3)}
\crefformat{figure}{Fig.~#2#1#3}
\crefrangeformat{figure}{Figs. #3#1#4--#5#2#6}
\crefmultiformat{figure}{Figs.~#2#1#3}{ and~#2#1#3}{, #2#1#3}{ and~#2#1#3}
\crefname{algocf}{Alg.}{Algs.}
\Crefname{algocf}{Algorithm}{Algorithms}
\crefformat{table}{Table~#2#1#3}
\crefrangeformat{table}{Tables~#3#1#4--#5#2#6}
\crefmultiformat{table}{Tables~#2#1#3}{ and~#2#1#3}{, #2#1#3}{ and~#2#1#3}

\usepackage{graphicx}
\usepackage{subcaption}
\usepackage{tikz, epstopdf, stfloats, bbding, capt-of}

\usepackage{algpseudocode}

\usepackage{fancyhdr}
\usepackage[most]{tcolorbox}
\definecolor{keyinsightbg}{HTML}{f7f7f7}     
\definecolor{keyinsightframe}{HTML}{1c2e4a}  
\definecolor{keyinsighttitle}{HTML}{000000}  

\usepackage{setspace}    
\newcounter{insightcounter}
\usepackage{palatino}  
\newtcolorbox{keyinsightbox}[1][]{
  enhanced,
  breakable,
  colback=keyinsightbg,       
  colframe=keyinsightframe,   
  coltitle=keyinsighttitle,   
  coltext=black,              
  fonttitle=\bfseries,
  width=\columnwidth,
  attach boxed title to top center={
    yshift=-\tcboxedtitleheight/2,
  },
  boxed title style={
    colback=white,            
    colframe=keyinsightframe, 
    coltitle=keyinsighttitle, 
    boxrule=1pt,
    arc=2pt,
    left=1pt,
    right=1pt,
    fonttitle=\bfseries,
  },
  before upper={
    \setstretch{1.2}          
    \itshape                  
    \begin{fontfamily}{ppl}\selectfont 
  },
  after upper={
    \end{fontfamily}
  },
  #1
}

%% file: custom.tex
\definecolor{Maroon}{RGB}{192,0,0}
\usepackage{wasysym}

\newcommand{\lightcomment}[1]{\textcolor{gray}{// #1}}

\newcommand{\sysname}{{Chiron}\xspace}



%% file: abstract.tex
\begin{abstract}
Large language model (LLM) serving is becoming an increasingly important workload for cloud providers.
Based on performance SLO requirements, LLM inference requests can be divided into 
\begin{enumerate*}[label=(\alph*)]
    \item interactive requests that have tight SLOs in the order of seconds, and
    \item batch requests that have relaxed SLO in the order of minutes to hours.
\end{enumerate*}
These SLOs can degrade based on the arrival rates, multiplexing, and configuration parameters, thus necessitating the use of resource autoscaling on serving instances and their batch sizes.
However, previous autoscalers for LLM serving do not consider request SLOs leading to unnecessary scaling and resource under-utilization.
To address these limitations, we introduce \sysname, an autoscaler that uses the idea of hierarchical backpressure estimated using queue size, utilization, and SLOs. 
Our experiments show that \sysname achieves up to 90\% higher SLO attainment and improves GPU efficiency by up to 70\% compared to existing solutions. 

\end{abstract}

%% file: 0000-intro_v2.tex
\section{Introduction}
\label{s:introduction}

\noindent \textbf{Background and Motivation.} 
Large language models (LLMs) such as OpenAI GPT-4, Google Gemini, and Meta Llama have enabled novel capabilities in a wide range of AI applications~\cite{compound-ai-blog,bommasani2021opportunities,wu2023autogen} such as chatbots and coding assistants.
Consequently, serving LLMs for enterprise and consumer applications with latency-oriented service-level objectives (SLOs) has become increasingly critical~\cite{jyothi2016morpheus,qiu2020firm,wang2024efficient}.
Based on their SLO requirements, requests can be divided into 
\begin{enumerate*}[label=(\alph*)]
    \item interactive requests that have tight SLOs in the order of seconds, and
    \item batch requests that have relaxed SLOs in the order of minutes to hours.
\end{enumerate*}

\begin{figure}[t]
    \centering
    \begin{minipage}[t]{0.238\textwidth}
        \centering
        \includegraphics{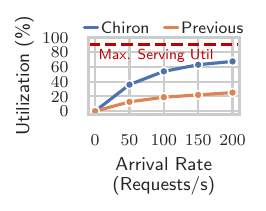}
        \label{fig:global_backpressure}
    \end{minipage}
    \hfill
    \begin{minipage}[t]{0.238\textwidth}
        \centering
        \includegraphics{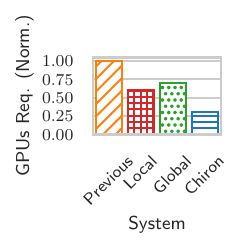}
        \label{fig:local_backpressure}
    \end{minipage}
    
    \caption{Previously proposed LLM serving systems overestimate backpressure leading to cluster-wide underutilization. (Left) Cluster-wide utilization when serving a mix of batch and interactive requests for Llama 8B and Llama 70B. (Right) GPUs required to serve the workload across various autoscalers. ``Local'' and ``Global'' are \sysname's autoscalers when used independently. }
    \label{fig:utilization_motivation}
\end{figure}

To serve these requests, LLM serving systems such as Llumnix~\cite{sun2024llumnix} and Ray Serve~\cite{rayserve} use an \emph{autoscaler} that scales resources based on the underlying backpressure generated from memory utilization and queuing metrics.
However, these autoscalers frequently overestimate backpressure at both local and global levels leading to excessive scale up actions and resource under-utilization as shown in Figure~\ref{fig:utilization_motivation} (Left).
Specifically, we find sub-optimality in the following two scenarios:
\begin{figure}
    \centering    \includegraphics{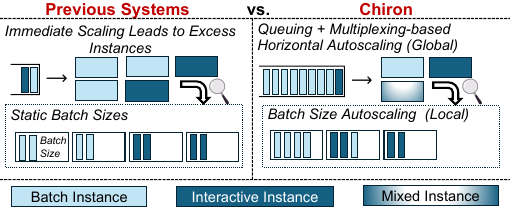}
    \caption{Illustration comparing \sysname with previous systems. \sysname uses less instances (five for \sysname vs. three for previous systems) because of (a) global autoscaling based on queuing and request multiplexing, and (b) local autoscaling based on dynamic batch sizes.}
    \label{fig:chiron_motivation}
\end{figure}

\emph{ At the local level, previous systems limit maximum batch sizes leading to sub-optimal serving throughput.}
Existing state-of-the-art serving systems such as vLLM use optimizations like continuous batching~\cite{yu2022orca} and PagedAttention~\cite{kwon2023efficient} to improve the serving throughput of LLM serving instances. 
While such optimizations significantly improve throughput, they also lead to increased request preemptions and inflate the inter-token latency for interactive requests beyond the SLO value.
To resolve this issue, in practice, operators limit the maximum batch size, diluting the benefits of PagedAttention.
However, when similar limitations are applied to batch requests, (shown in Figure~\ref{fig:chiron_motivation} (Left)), it can lead to suboptimal utilization and reduced throughput.

\emph{ At the global level, previous systems excessively scale up the number of instances with the arrival of batch requests.}
Previous systems do not distinguish between request with varying SLOs when performing scaling actions.
Consequently, as batch requests arrive there is an immediate scale up of instances as shown in Figure~\ref{fig:chiron_motivation} (Left).
Instead, batch requests can be maintained in a queue and scaling action can be delayed to allow for multiplexing with interactive requests and improve resource utilization ( as shown in Figure~\ref{fig:chiron_motivation} (Right)).

\noindent \textbf{Our Work.}
To address the above limitations, we propose \sysname, a pluggable autoscaler that works with exisiting LLM serving systems.
\sysname uses the idea of \textbf{hierarchical backpressure} to determine scaling actions at the  local and global serving instances.

At the local instance level, batch size controls number of requests served concurrently.
If batch size increases throughput is higher, albeit at the cost of latency.
Previous attempts at batch size optimization such as Andes~\cite{liu2024andes} and Vidur~\cite{agrawal2024vidur} typically rely on offline profiling.
However, such an approach is increasingly infeasible with the exponential configuration space for optimizations such as speculative decoding, prefix caching, chunked prefill, etc.
Instead, \sysname uses a reactive autoscaling approach based on local backpressure generated from a combined latency throughput metric.
If latency SLO violation or throughput degradation is observed, local backpressure increases and \sysname reduces the batch size to eliminate preemptions and meet the SLO (and vice-versa if SLOs are already met).

At the global cluster level, the number of serving instances determine the overall throughput and latency serving capabilities of the system.
An ideal autoscaling policy would dynamically change the number of instances to match the incoming request workload.
However, with LLM serving workloads, unlike previous cloud workloads, time to bring up new instances is much higher due to large model size.
Therefore, the system would always have to be over-provisioned to ensure that SLOs for interactive requests are not missed during bursty arrivals.
As such over-provisioning is necessary, it leads to spare capacity in the system that can be leveraged by batch requests to further improve utilization and throughput.
To decide whether the spare capacity is sufficient to meet SLOs and scale instances accordingly, \sysname uses request waiting time estimation from QLM~\cite{patke2024one}.
Additionally, \sysname uses requests groups from SHEPHERD~\cite{zhang2023shepherd} to minimize unnecessary scaling actions.

\noindent \textbf{Results.}
We demonstrate \sysname on vLLM~\cite{kwon2023efficient} as the backend LLM-serving system and GPU clusters with NVIDIA A100 GPUs.
We adopt workloads from a real-world LLM dataset: ShareGPT~\cite{shareGPT} using setups derived from our production requirements.
Our experiments demonstrate the following major improvements with \sysname:
\begin{enumerate}[wide=0pt,label=(\arabic*)]
    \item \emph{SLO Attainment:} Depending on the arrival rate, \sysname achieves up to 90\% higher SLO attainment compared to previous LLM serving systems like Llumnix.
    \item \emph{GPU Efficiency Improvements:} \sysname improves the request throughput up to 300\% compared to previous systems. Such throughput increase leads to GPU savings up to 70\% as shown in Figure~\ref{fig:utilization_motivation} (Right).
    \item \emph{Ablation Study and Robustness Analysis:} We demonstrate that using both local and global backpressure contributes to SLO attainment and throughput improvement.
    Additionally, we present robustness analysis with varying SLO values and bursty arrival patterns.
\end{enumerate}
 

%% file: 0001-background.tex
\section{Background}
\label{s:background}

\subsection{LLM Inference}
\noindent
\textbf{Inference Primer.}
An inference process starts from a request (prompt) with a list of input tokens from which 
the LLM generates a list of output tokens .
Due to the \textit{autoregressive} pattern, the LLM can only generate new tokens one by one, and the generation process of each new token depends on all the previous tokens in that sequence, specifically their key and value vectors.
In this sequential generation process, the key and value vectors of existing tokens are cached for generating future tokens, known as \textit{KV cache}.

\noindent
\textbf{Continuous Batching.}
During LLM inference, the decoding stage is memory-bound, as loading model weights from memory takes longer than computation.
Therefore, state-of-the-art LLM serving systems like vLLM~\cite{kwon2023efficient}, Orca~\cite{yu2022orca}, Tensor-RT~\cite{tensorrt} and TGI~\cite{tgi} employ continuous batching with iterative scheduling to enable dynamic addition of requests to a batch as soon as others have finished generation.

\noindent
\textbf{PagedAttention.}
Static allocation of the KV cache can result in significant memory waste as the KV cache grows dynamically during the decoding stage.
PagedAttention~\cite{kwon2023efficient} introduces the idea of managing the KV cache, like OS memory, via pages and enabling dynamic allocation.
Such dynamic allocation prevents fragmentation and enables nearly 100\% utilization of GPU memory and furthers throughput improvement when combined with continuous batching.

\noindent \textbf{Workload Categories.}
LLM serving requests can be divided into two types:
\begin{enumerate*}[label=(\alph*)]
\item \emph{Interactive: } User-facing and agentic applications like chatbots and coding assistants require immediate responses.
\item \emph{Batch: } Background applications such as document processing, log processing, and synthetic data generation allow for longer completion times, ranging from minutes to hours.
\end{enumerate*}

\begin{figure}
    \centering
    \begin{subfigure}[t]{0.22\textwidth}
        \centering
        \includegraphics[width=\textwidth]{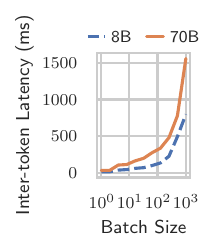}
        \label{fig:itl_vs_batch}
    \end{subfigure}
    \hfill
    \begin{subfigure}[t]{0.22\textwidth}
        \centering
        \includegraphics[width=\textwidth]{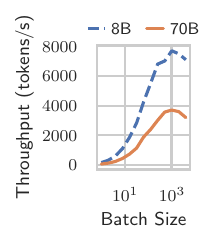}
        \label{fig:tput_vs_batch}
    \end{subfigure}
    \caption{Variation in inter-token latency and token throughput with increasing batch size.}
    \label{fig:comparison}
\end{figure}

\subsection{Definitions}

\begin{definition}
    \emph{SLO:} 
    Each LLM serving request has two key latency requirements:
    \begin{enumerate*}[label=(\alph*)]
        \item \emph{time to first token (TTFT):} the time required to complete the prefill step and generate the first token, and
        \item \emph{inter-token latency (ITL): } the time required to generate each subsequent token in the decode phase.
    \end{enumerate*}
    These two latency requirements together form the service-level objective (SLO) for the request.
\end{definition}

\begin{definition}
\label{def:serving-instance}
    \emph{LLM Serving Instance:} An LLM serving system is capable of hosting LLM models by providing the necessary infrastructure and resources to load the models into memory and respond to requests.
    \sysname is compatible with existing LLM serving systems such as vLLM~\cite{kwon2023efficient} and TGI~\cite{tgi}.
    An LLM serving instance is composed of the LLM serving system and an LLM model that is being served.
\end{definition}

\begin{definition}
    \label{def:autoscaler}
    \emph{Autoscaler: } An autoscaler provisions resources based on incoming request arrival patterns.
    An ideal autoscaler
    \begin{enumerate*}[label=(\alph*)]
        \item maximizes serving throughput while maintaining request SLOs,
        \item quickly converges to optimal resource levels, and
        \item avoids unnecessary scaling actions (referred to as hysteresis).
    \end{enumerate*}
    Note that such an autoscaler also ensures high system utilization and reduces total resource requirement.
\end{definition}

\subsection{Motivation and Characterization}
\label{ss:characterization}

To design an SLO-aware autoscaler for LLM workloads, we investigate the following scenarios:
\begin{enumerate*}[label=(\alph*)]
    \item Request arrival patterns in a production cluster,
    \item Limitations of existing autoscalers with varying SLO requirements, and 
    \item Throughput versus latency trade-off space in LLM serving systems.
\end{enumerate*}
Besides data from the internal production cluster, we use state of the art LLM serving system, vLLM, to investigate these scenarios using the ShareGPT~\cite{shareGPT}  dataset.  


\begin{figure*}[!ht]
    \centering
    \begin{minipage}[t]{.32\textwidth}
        \centering    \includegraphics{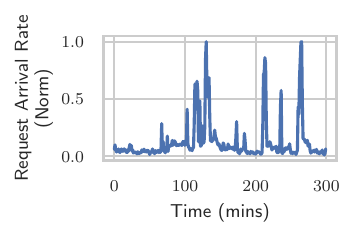}
        \caption{Request arrival spikes in a production serving cluster over a 5 hour duration.}
        \label{fig:bursty_arrivals}
    \end{minipage}%
    \hfill
    \centering    
    \begin{minipage}[t]{.32\textwidth}
        \centering
        \includegraphics{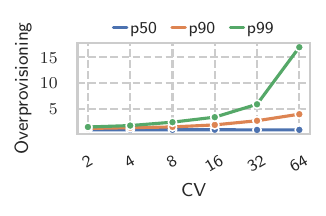}
        \caption{Over-provisioning required for varying burstiness.}
        \label{fig:oveprovisioning_gamma}
    \end{minipage}%
    \hfill
    \centering    
    \begin{minipage}[t]{.32\textwidth}
    \centering
    \includegraphics{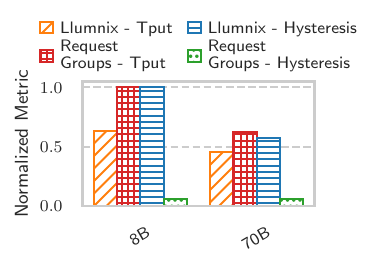}
    \caption{Request groups can prevent hysteresis associated with autoscaling actions.}
    \label{fig:hysteresis_prevention}
    \end{minipage}%
    \hfill
\end{figure*}

\noindent \textbf{\textit{Burstiness in request arrivals leads to SLO violations in production clusters.}}
An optimal autoscaling policy scales instances up and down to match the incoming request arrival rate while minimizing any under-utilized capacity.
However, for LLM workloads, the time to scale up new instances is significant due to the large model sizes.
Even with model loading optimizations~\cite{fu2024serverlessllm}, the model load time varies between 15 seconds and one minute.
Due to these long model load times, new instances cannot be brought online quickly enough when the system is at capacity and the request arrival rate increases (i.e., a request spike occurs), leading to SLO violations.

To quantify this observation, we define a request spike as the ratio of the request arrival rate between two consecutive time intervals, where the interval length corresponds to the model load time.
If an arrival spike is greater than one, and the system is at capacity (i.e., no GPU memory is available to process additional requests), an SLO violation will occur.
We observe that such large arrival rate spikes are common in our internal production cluster traces, as shown in Figure~\ref{fig:bursty_arrivals}.
Over a two-month period, the p90 and p99 values of the arrival spikes are 1.6 and 3, respectively.

Additionally, we created a synthetic workload with Gamma arrival rates to simulate increasing burstiness levels by varying the coefficient of variance (CV) parameter~\cite{zheng2022alpa}. Figure~\ref{fig:oveprovisioning_gamma} shows the necessary over-provisioning to meet SLOs across different percentiles. 
As burstiness increases, requests spikes increase and additional provisioning is necessary.

\begin{keyinsightbox}
\begin{itemize}
    \item Resource over-provisioning is necessary to serve interactive requests and such over-provisioning is proportional to burstiness of request arrivals.
    \item \textbf{Design Consequence 1: } \sysname uses extra capacity to serve batch requests via a global autoscaler.

\end{itemize}
\end{keyinsightbox}

\textbf{\textit{Inter-token latency and throughput can vary significantly with changes in batch size.}}
A larger batch size results in longer self-attention computation times and more preemptions, which can increase inter-token latency and potentially cause SLO violations.
To address this, cloud providers typically impose a maximum batch size limit to meet inter-token latency SLOs.

However, larger batch sizes can also be advantageous, as they allow more requests to be processed simultaneously, improving throughput for batch requests.

Figure~\ref{fig:comparison} shows the variation in throughput and latency for two models, Llama-8B and Llama-70B. 
We observe that as batch size increases, both inter-token latency and throughput increase.

\begin{keyinsightbox}
\begin{itemize}
    \item LLM serving instances that serve interactive requests should use smaller batch sizes to keep inter-token latency within the SLO, while instances that serve batch requests can use a larger batch size to optimize throughput.
    \item For instances that process both interactive and batch requests, the interactive request SLO can be used to determine batch size.
    \item \textbf{Design Consequence 2:} \sysname uses three kinds of serving instances: interactive, batch, and mixed.
    Batch size across these instances is decided by a local autoscaler.
\end{itemize}
\end{keyinsightbox}

\noindent \textbf{\textit{Immediate autoscaling can lead to hysteresis and reduced serving throughput.}}
Interactive requests must be served immediately upon arrival, whereas batch requests can be queued because they do not require immediate processing due to their higher SLO values.
Such queuing allows batch requests to opportunistically utilize the over-provisioned capacity reserved for interactive requests (i.e., through request multiplexing).
We evaluate this benefit in Section~\ref{s:results}.
To maximize the benefits of such multiplexing, the optimal autoscaling strategy would involve adding instances only when spare capacity is insufficient and a batch request is approaching its deadline.

However, previous autoscalers such as Llumnix add instances immediately upon request arrival and remove them upon request completion leading to significant churn, commonly referred to as \emph{hysteresis}.
Specifically, we define hysteresis as the ratio of total scaling actions (including scale up and scale down) to the total scale up actions across the experiment duration.
To minimize hysteresis, requests with similar SLO deadlines can be grouped and processed together.
By serving requests in groups, frequent scaling actions are reduced, resulting in an improvement in average serving throughput.
For instance, Figure~\ref{fig:hysteresis_prevention} demonstrates that hysteresis can reduce by 20$\times$ and throughput can improve by 2.5$\times$ when requests are processed in groups.

\begin{keyinsightbox}
\begin{itemize}
    \item \textbf{Design Consequence 3: } \sysname uses request groups to minimize hysteresis associated with repeating autoscaling actions and improve serving throughput.
\end{itemize}
\end{keyinsightbox}

%% file: 0003-rpms.tex
\section{\sysname Design and Implementation}
\label{s:design}

\begin{figure}[!ht]
\centering
    \includegraphics{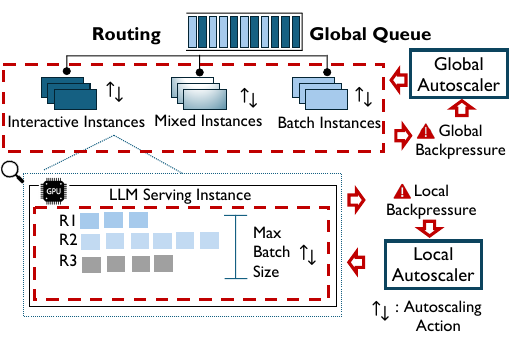}
    \caption{Overview of \sysname.}
    \label{fig:chiron_overview}
\end{figure}

Figure~\ref{fig:chiron_overview} provides an overview of \sysname. 
\sysname follows a hierarchical autoscaling design to meet TTFT and ITL SLOs while maximizing throughput by:
\begin{enumerate*}[label=(\alph*)]
\item  scaling the batch size of an individual instance via \emph{Local Autoscaler}, and \item  scaling the interactive, mixed and batch instances (i.e., horizontal scaling) via \emph{Global Autoscaler}
\end{enumerate*}

\noindent \textbf{Lifecycle of a Request.}
All incoming requests are enqueued into a \emph{global queue} and then served by the underlying LLM serving instances.
Each serving instance is classified into one of three categories:
\begin{enumerate*}[label=(\alph*)]
    \item \emph{Interactive Instances} serve interactive requests only,
    \item \emph{Mixed Instances} serve both interactive and batch requests, and
    \item \emph{Batch Instances} serve batch requests only.
\end{enumerate*}
Each request is preferentially routed to it's own instance type (i.e., interactive requests to interactive instances and batch requests to batch instances) leading to non-uniform routing of requests in \sysname.
If capacity is unavailable on its own instance type, they are routed to the mixed instances. 






\noindent \textbf{Request Multiplexing with Mixed Instances.}
Mixed instances enable multiplexing between interactive and batch requests and drive up overall cluster utilization.
For interactive requests, the mixed instances can handle unpredictable spikes in request arrivals.
For batch requests, the mixed instances provides additional running capacity when sufficient interactive requests are not present.
To enable such multiplexing between interactive and batch requests while ensuring immediate execution of interactive requests, mixed instances are preemptible, i.e., interactive requests can evict out batch requests and send them back into the global queue.
To prevent throughput drop from such eviction, we enable fast restart by saving the KV cache by migrating it to CPU memory.

%% file: 0003-local.tex
\section{Local Autoscaler}
\label{s:local_autoscaler}

The local autoscaler uses an online algorithm to determine the optimal maximum batch size for each instance using a \emph{local backpressure} metric, that accounts for both latency and throughput degradation.





\textbf{Why use an online algorithm?}
In production settings, each serving instance can be configured with multiple optimizations, such as speculative decoding, prefix caching, and chunked prefill.
These configurations impact latency and throughput of the serving instance.
However, the usual practice of estimating the optimal batch size per instance through offline profiling is prohibitively expensive due to exponential configuration space. 

\subsection{Local Backpressure}
Increase in latency and throughput degradation at the local instance-level leads to local backpressure, that we define as follows:
\begin{itemize}[wide=0pt]
\item  \emph{Latency-based Backpressure (LBP): }
Empirically, we observe that the increasing the batch size increases the inter-token latency (ITL) due to increased request preemptions and higher self-attention computation costs.
Such increase in inter-token latency can lead to SLO violations.
To account for the ITL inflation, we define the latency-based backpressure (LBP) metric as the ratio between observed ITL and the ITL SLO value.
When LBP is greater than one, ITL SLO violations are observed and the batch size should be decreased.\\

\item \emph{Throughput-based Backpressure (TBP): }
Increase in batch size can also increase throughput as more requests are processed simultaneously.
\textit{However, surprisingly, we observe that increasing the batch size beyond an inflection point results in a decrease in throughput due to increased preemptions and higher self-attention costs (similar to inter-token latency increase).}
Figure~\ref{fig:comparison} shows the inflection points for Llama 8B and 70B. 
To account for this drop in throughput, we define throughput-based backpressure (TBP) as the ratio between previously observed and current throughput.
If TBP is greater than one, no throughput gain is observed from increasing the batch size and it should be decreased.
\end{itemize}

\subsection{Batch Size Autoscaling}

Algorithm~\ref{algo:local_autoscaling} describes the overall autoscaling algorithm that varies the batch size based on local backpressure.
If local backpressure is greater than one, the batch size is divided by half.
Otherwise, the batch size is increased proportionally with exponentially weighted moving average (EWMA) to obtain throughput benefits.
EWMA and proportional scaling allows for convergence to the optimal batch size value\footnote{EWMA is widely used in resource control algorithms such as congestion control~\cite{jacobson1988congestion}, operating system load estimation~\cite{bovet2005understanding}, and database query optimization~\cite{boncz1999database}.}.
As the backpressure approaches one, the algorithm slows down the increase in batch size values.
Note that the ITL SLO for the instance is the smallest ITL SLO amongst all requests executing on that instance. 


\begin{algorithm}[t]
\caption{Batch Size Autoscaling Algorithm}\label{alg:batch_size_backpressure}
\begin{algorithmic}[1]
\raggedright
\State \textbf{Input:} ITL SLO, EWMA Smoothing factor ($\alpha = 0.5$)

\State \lightcomment{Update loop for the autoscaler}
\For{each change in the GPU running queue}

\State \lightcomment{Compute local backpressure}
    \State LBP = $\gets$ ITL/{ITL SLO}
    \State TBP $\gets$ $\text{Throughput}_{prev}$ / $\text{Throughput}_{curr}$
    \State Local Backpressure $\gets \max(\text{LBP},\text{TBP})$
    \If{Local Backpressure $< 1$}

        \State \lightcomment{Scale up batch size with EWMA}
        \State \text{Max Batch Size} $\gets \alpha \times 1/\text{Local Backpressure}$ 
        
        \State $\times \text{Max Batch Size} + (1 - \alpha) \times \text{Max Batch Size}$
    \Else
    
        \State \lightcomment{Scale down batch size}
        \State $\text{Max Batch Size} \gets \text{Max Batch Size} / 2$
    \EndIf
\EndFor
\end{algorithmic}
\label{algo:local_autoscaling}
\end{algorithm}

%% file: 0003-global.tex


\section{Global Autoscaler}
\label{s:global_autoscaler}

The global autoscaler scales LLM serving instances based on the incoming request using a \emph{global backpressure} metric that accounts for instance usage and queue formation.

\subsection{Global Backpressure}
Global backpressure can arise from both increased instance utilization and request queue formation as described below:
\begin{itemize}[wide=0pt]
    \item \emph{Interactive Backpressure (IBP): } Interactive instances need to be overprovisioned to handle any unpredictable spikes in request arrivals, as shown earlier in Section~\ref{ss:characterization}.
    In the case of \sysname, mixed instances represent such over-provisioned capacity and request spikes are routed to the mixed instances, leading to increased backpressure as their usage increases.
    Therefore, we define interactive backpressure (IBP) as
    the ratio of instances running interactive requests to the total mixed and interactive instances.
    As IBP increases new mixed and interactive instances need to be added to maintain the level of over-provisioning.\\
    \item \emph{Batch Backpressure (BBP): } Batch requests can utilize the spare capacity available on the mixed instances.
    However, such spare capacity may not be sufficient to complete all batch requests and they would remain in the queue.
    Increase in queue by itself is not a problem if requests are not close to their TTFT SLO-based deadline.
    Hence, we define batch backpressure (BBP) as number of batch requests in the global queue that are close to their TTFT SLO (formally defined in Section~\ref{ss:batch_autoscaling}).
    As BBP increases, the queue waiting time can exceed the TTFT SLO-based deadline and new batch instances need to be added.
\end{itemize}

\subsection{Interactive Autoscaling}

The interactive autoscaler attempts to maintain a constant level of over-provisioning (i.e. the ratio of instances running interactive requests to the total mixed and interactive instances) equal to $\Theta$.
As requests arrive, the level of over-provisioning varies leading the IBP (i.e., the current over-provisioning level) to change.
If IBP exceeds $\Theta$, interactive and mixed instances are added to maintain over-provisioning.
Conversely, if IBP decreases below $\Theta$, mixed and interactive instances are removed.
The specific value of $\Theta$ is chosen based on historic arrival patterns in the system.
For example, if the tail request arrival spike is three times the average arrival rate, $\Theta$~\footnote{In practice, to minimize hysteresis (i.e. constant addition/retirement of instances), we maintain IBP in a range equal to [$\Theta-\delta,\Theta+\delta$]} is set to $1/3$.

\subsection{Batch Instance Autoscaling}
\label{ss:batch_autoscaling}

The batch instance autoscaler scales batch instances based on the requests in the queue using BBP as shown in Algorithm~\ref{algo:batch_autoscaling}.\footnote{Note that interactive requests follow a zero-queuing approach to minimize TTFT.}

\emph{Request Group Creation: }
BBP uses the idea of request groups, which are created by clustering requests in the queue with similar TTFT SLO values~\cite{macqueen1967some}.
Since requests within a request group have similar TTFT SLO requirements, \sysname treats the ordering of the requests within a group using a first-come-first-serve (FCFS) policy.
These request groups are executed together, thus minimizing hysteresis as discussed in Section~\ref{ss:characterization}.

\begin{algorithm}[t]
\caption{Batch Instance Autoscaling Algorithm}
\label{algo:batch_autoscaling}
\begin{algorithmic}[1]
\State \textbf{Input:} Batch Queue, TTFT SLO
\State \textbf{Initialize:} $\text{dispatch instances} \gets 0$
\State \lightcomment{Keep adding instances until backpressure is 0}
\While{$BBP > 0$}
    \State $BBP \gets 0$
    \State $\text{dispatch instances} \gets \text{dispatch instances} + 1$
    \State \lightcomment{Calculate batch backpressure from queue state}
    \For{each request group $g$ in Batch Queue}
        \State $W_g \gets \text{EstimateWaitingTime}(g)$
        \If{$W_g > \text{TTFT SLO}$}
            \State $BBP \gets BBP + 1$
        \EndIf
    \EndFor
\EndWhile
\State \lightcomment{Add minimum calculated instances}
\State Add batch instances$(\text{dispatch instances})$
\State \lightcomment{Retire all instances if no batch requests are served}
\If{no active requests}
    \State Remove all batch instances
\EndIf
\end{algorithmic}
\end{algorithm}

\emph{Estimating Batch Backpressure (BBP):}
\sysname estimates the queue waiting time for each request group to determine BBP.
If requests with waiting time is close to the TTFT SLO deadline, the BBP increases (and vice-versa).
The waiting time estimates are generated with a statistical approach as described in QLM~\cite{patke2024one}.
We describe the estimation approach below for completeness.
\sysname consider the token generation throughput ($\Theta$) to be constant throughout the token generation process due to statistical averaging effects of continuous batching.
Therefore, the total waiting time for a single request can be represented by Equation~\ref{eq:base_equation} by dividing the number of tokens ahead ($\sum_{i=1}^{q-1} O_i$) in the queue by the token generation throughput ($\Theta$) where $i$ denotes each of the $q-1$ requests in the queue ahead of the request we model.
\begin{align}
W_q = \sum_{i=1}^{q-1} \frac{O_i}{\Theta}
\label{eq:base_equation}
\end{align}

\noindent
Note that we do not know the number of output tokens ahead of time (that requires the knowledge of the output sequence for all requests in the waiting queue), so we model them as a distribution with the mean $\mu_o$ and standard deviation $\sigma_o$ fitted from previous requests arriving at the instances.
As $q$ increases, the Central Limit Theorem (CLT) applies and the assumption of Normal distribution is accurate for any underlying request output token distribution.

Finally, the BBP is simply the number of request groups with waiting time that exceeds the TTFT SLO.

\begin{align}
    BBP = \sum_{i=1}^{n} \mathbb{I}(W_i > \text{TTFT SLO})
\end{align}

\emph{Autoscaling Algorithm: }
The final autoscaling algorithm shown in Algorithm~\ref{algo:batch_autoscaling} adds the minimum instances required to make the value of BBP equal to zero.
If BBP is zero and there are no active requests being served on the batch instances, then all batch instances are retired.

%% file: 0006-results.tex
\section{Evaluation}
\label{s:results}



We evaluate \sysname on the following dimensions:
\begin{enumerate}[label=(\alph*)]
    \item SLO attainment and throughput improvements for an interactive workload (\cref{ss:single_model_eval}),
    \item SLO attainment and throughput improvements for a combined interactive and batch workload (\cref{ss:multi-model-eval}),
    \item Time for autoscaler to converge,
    \item Robustness analysis including accuracy of queue waiting time estimator, use of serving optimizations, variable SLO values, and bursty arrival patterns, and
    \item Ablation study showing benefits of the local and global autoscaler.
\end{enumerate}

\noindent \textbf{Experiment Setup.}
We evaluate \sysname on two open-source LLMs: 
\begin{enumerate*}[label=(\alph*)]
    \item Meta Llama 3.1 8B~\cite{touvron2023llama} and
    \item Meta Llama 3.1 70B~\cite{touvron2023llama}
\end{enumerate*}.
Both models were configured with additional optimizations such as prefix caching and speculative decoding.
For the speculative decoding configuration, Llama 70B was configured with Llama 8B as the draft model.
We evaluate \sysname on an elastic cloud, however we cap the total GPUs to 50 NVIDIA A100 (80 GB memory).
We compare \sysname against the following two baselines:
\begin{enumerate*}[label=(\alph*)]
    \item \emph{Llumnix}: A state-of-the-art LLM orchestration and serving system based on vLLM.
    The autoscaler in Llumnix keeps average token utilization across all instances between a configurable threshold range by adding and removing serving instances.
    For the base Llumnix version, we use the configuration that maximizes SLO satisfaction across all workloads.
    \item \emph{Llumnix (tuned)}: In this version of Llumnix, we perform a parameter sweep per-workload to find the configuration that maximizes SLO attainment and throughput.
\end{enumerate*}


\begin{figure}[!h]
\centering
    \includegraphics{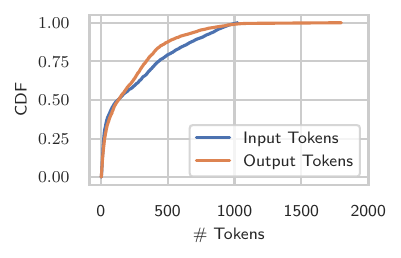}
    \caption{Token distribution in the ShareGPT dataset.}
    \label{fig:token_distribution}
\end{figure}

\noindent
\textbf{Workloads.}
We create our experimental workloads from the requirements of a production cloud service provider except for request arrival rates due to confidentiality reasons.
Request arrivals are modeled with a Poisson distribution and backpressure is created by varying the arrival rates.
Each workload trace uses 3,500 requests from the ShareGPT~\cite{shareGPT} dataset with input/output token distribution as shown in Figure~\ref{fig:token_distribution}.

Each request is one of two types
\begin{enumerate*}[label=(\alph*)]
    \item \emph{Interactive:} These requests represent chat applications with a TTFT SLO of ten seconds and ITL SLO of 200 ms based on average human reading speed,
    \item \emph{Batch:} These requests represent document processing or data generation tasks with a TTFT SLO of one hour and ITL SLO of two seconds.
\end{enumerate*}
While our choice of SLO values is motivated by production requirements, we also provide robustness analysis for alternative SLO values.

We create our workloads using a combination of these requests as follows:
\begin{enumerate*}[label=\textbf{[$\mathbf{W_\Alph*}$]}]
    \item \textbf{\textit{Interactive-only Workload}} which consists of Interactive requests for small model, large model and a mixed configuration that uses both models.
    In the mixed configuration, all interactive requests are equally divided between the small and large models.
    \item \textbf{\textit{Interactive and Batch Workload}} which consists of interactive and batch requests for the small model, large model, and a mixed configuration.
    In the mixed configuration, both interactive and batch requests are equally divided between the small and large model.
\end{enumerate*}

\begin{figure}[!t]
    \includegraphics[]{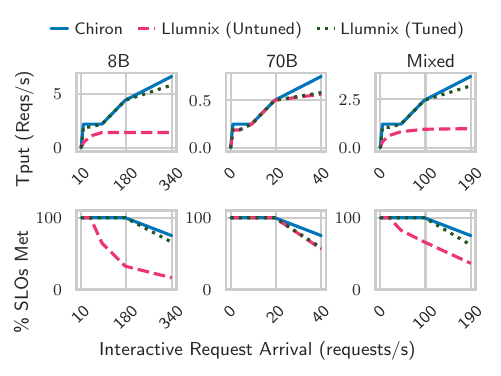}
    \caption{$W_A$: Interactive workload with varying arrival rates for small model, large model, and mixed model configurations.}
    \label{fig:interactive}
\end{figure}


\subsection{Interactive Autoscaling}
\label{ss:single_model_eval}

To understand the efficacy of \sysname's interactive autoscaling, we run workload $W_A$ with varying interactive arrival rates and evaluate the impact on throughput and SLO satisfaction metrics.
Figure~\ref{fig:interactive} shows the average per-instance throughput and SLOs met across all autoscalers for the small, large, and mixed model configurations.

\noindent
\textbf{Request Throughput.}
For all configurations, we find that \sysname has equal or higher per-instance throughput compared to both versions of Llumnix.
Specifically, \sysname performs better because:
\begin{enumerate*}[label=(\alph*)]
    \item Compared to Llumnix (Untuned), \sysname is able to adapt batch sizes across different models. 
    For example, the small model can use a large batch size with higher throughput, whereas the large model would use a smaller batch size to maintain the same inter-token latency SLO.
    The local autoscaler in \sysname is able to consider these differences and maintains a 5$\times$ higher batch size for the small model compared to the large one.
    \item Compared to Llumnix (Tuned), \sysname adjusts batch size dynamically across time based on the token distributions.
    For example, when multiple short requests are processed, \sysname adapts to a higher batch size, and vice-versa.
    While such dynamic adaptation to token distribution is not as significant as changing models, it still leads to a 10\% throughput improvement.
    For example, this occurs at an arrival of 340 requests/s for the small model and 40 requests/s for the large model.
\end{enumerate*}

\noindent
\textbf{SLO Attainment.}
\sysname is able to achieve better SLO satisfaction than Llumnix (Tuned and Untuned) due to the throughput gains.
As Llumnix throughput is smaller than \sysname, it autoscales faster and consumes all available GPUs.
Any further increase in request arrival rate beyond this point leads to SLO violations.
For example, in the mixed model configuration at 40 requests/s, Llumnix (Untuned) exhausts all available GPUs leading to SLO violations.
As request arrival rates increase further, \sysname also utilizes all GPUs leading to SLO violations beyond a threshold of 100 requests/s.

\begin{figure}[!t]
    \includegraphics[]{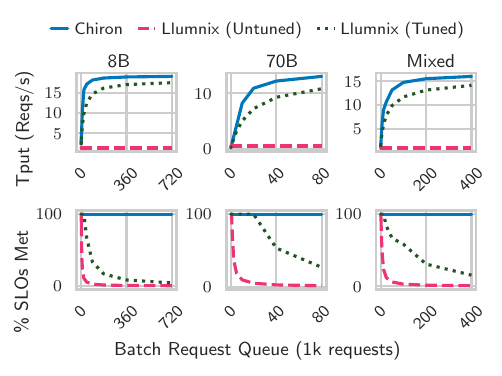}
    \caption{$W_B$: Interactive and Batch workload with varying batch request queue and fixed interactive arrival rate for small model, large model, and mixed model configurations.}
    \label{fig:batch}
\end{figure}

\subsection{Batch Autoscaling}
\label{ss:multi-model-eval}
To understand the efficacy of \sysname's batch autoscaler, we run workload $W_B$ with varying batch request queues and evaluate the impact on request throughput and SLO satisfaction.
In addition to the batch requests, we also serve interactive requests with constant arrival rate of 50 requests/s and 10 requests/s for 8B and 70B models respectively.
Figure~\ref{fig:batch} shows the average per-instance throughput and SLOs met across all autoscalers for the small, large, and mixed model configurations.

\noindent
\textbf{Request Throughput.}
Similar to $W_A$, we find that \sysname has higher per-instance throughput with $W_B$ across all configurations compared to other autoscalers.
Specifically, we find that
\begin{enumerate*}[label=(\alph*)]
    \item Compared to Llumnix (Untuned), \sysname uses a higher batch size. As batch requests have a relaxed ITL SLO, \sysname's autoscaler sets a nearly 50$\times$ higher batch size (between 2048--4096) resulting in substantial throughput improvement,
    \item Compared to Llumnix (Tuned), \sysname can multiplex batch and interactive requests on the mixed instances, thus increasing the utilization of the over-provisioned capacity on the system.
\end{enumerate*}
Due to the above mentioned optimizations, \sysname is able to handle a batch request queue of 700k requests and 80k requests for the small and large model respectively.




\noindent
\textbf{SLO Attainment.}
\sysname also achieves higher SLO attainment compared to other autoscalers because
\begin{enumerate*}[label=(\alph*)]
    \item \sysname's higher throughput enables a higher number of requests to be processed with limited capacity, and
    \item Recall from Algorithm~\ref{algo:batch_autoscaling}, at higher batch request queue sizes, \sysname adds multiple batch instances at once based on the queue waiting time estimation.
    In contrast, Llumnix increases capacity gradually one instance at a time resulting in longer warm-up times and further reduced throughput.
\end{enumerate*}

\begin{figure*}[!t]
    \centering    
    \begin{minipage}[t]{.24\textwidth}
        \centering
        \includegraphics{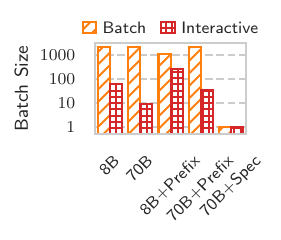}
        \caption{Batch size variation across different configurations for interactive requests.}
        \label{fig:batch_size_variation}
    \end{minipage}
    \hfill
    \centering    
    \begin{minipage}[t]{.24\textwidth}
        \centering
        \includegraphics{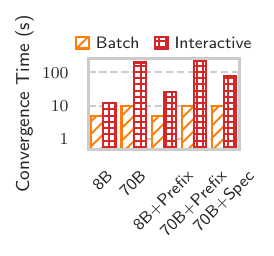}
        \caption{Varying convergence time across different configurations.}
        \label{fig:convergence_time}
    \end{minipage}
    \hfill
    \centering 
    \begin{minipage}[t]{.25\textwidth}
        \centering
        \includegraphics{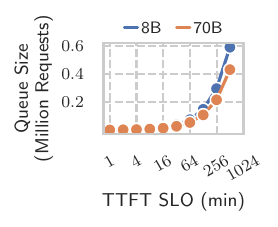}
        \caption{Queue sizes maintained for varying batch SLO.}
        \label{fig:queue_sizes}
    \end{minipage}%
    \hfill
    \centering
    \begin{minipage}[t]{.23\textwidth}
        \centering
        \includegraphics{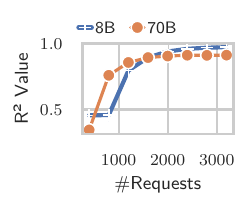}
        \caption{Accuracy of queue waiting time estimation.}
        \label{fig:rwt_estimator}
    \end{minipage}%
    \hfill
\end{figure*}

\begin{figure}[t]
    \begin{minipage}[bt]{.20\textwidth}
        \centering
        \includegraphics{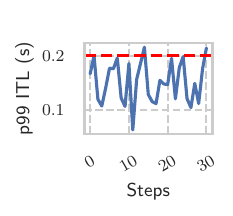}
        \caption{ITL variation across multiple local autoscaling steps.}
        \label{fig:itl_vs_steps}
    \end{minipage}
    \hfill
    \centering    
    \begin{minipage}[bt]{.26\textwidth}
    \centering
    \resizebox{1.8in}{!}{
    \begin{large}
    \begin{tabular}{|@{\hskip 2pt}c@{\hskip 2pt}|@{\hskip 2pt}c@{\hskip 2pt}|@{\hskip 2pt}c@{\hskip 2pt}|@{\hskip 2pt}c@{\hskip 2pt}|}
    \hline
    \textbf{\makecell{ITL \\ SLO (s)}} & \textbf{\makecell{\% SLOs \\ Met}} & \textbf{\makecell{Throughput \\ (Requests/s)}} & \textbf{\makecell{\#GPUs \\ Required}} \\[5pt]
    \hline
    0.1 & 99.3 & 1.1 & 100\% \\[5pt]
    0.2 & 99.7 & 2.8 & 39\% \\[5pt]
    1   & 100  & 9 &  12\%\\[5pt]
    10  & 100  & 14 & 8\% \\[5pt]
    100 & 100  & 16 & 7\% \\[5pt]
    \hline
    \end{tabular}
    \end{large}
    }
    \caption{ITL SLO satisfaction for varying values, with corresponding throughput and GPUs required for Llama 70B.}
    \label{table:itl_values}
\end{minipage}
\hfill
\end{figure}

\begin{figure}[ht]
    \begin{minipage}[t]{.23\textwidth}
        \centering
        \includegraphics{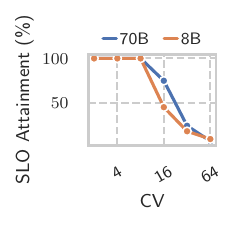}
        \caption{SLO satisfaction with varying burstiness in request arrivals.}
        \label{fig:bursty_slos}
    \end{minipage}
    \hfill
    \centering    
    \begin{minipage}[t]{.23\textwidth}
    \includegraphics{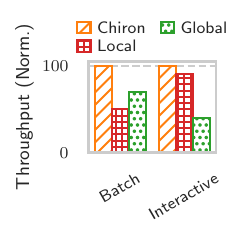}
    \caption{Ablation study for local and global autoscaler.}
    \label{fig:ablation}
    \end{minipage}
\hfill
\end{figure}

\subsection{\sysname Robustness Analysis}
\label{ss:robustness} 

\noindent
\textbf{Convergence Analysis.}
\sysname's local autoscaler dynamically adapts the batch size based on the underlying configurations to meet the ITL SLO while maximizing serving throughput as shown in Figure~\ref{fig:batch_size_variation}.
Specifically, we observe the following trends:
\begin{enumerate*}[label=(\alph*)]
    \item If prefix caching is enabled the final converged batch size reduces as a larger KV cache is loaded at the beginning leading to higher memory utilization.
    Such high memory utilization causes request preemptions and inflates the inter-token latency.
    To mitigate this effect of preemptions, the batch size has to be reduced.
    \item Speculative decoding also prefers smaller batch sizes to minimize interference with the draft model execution.
\end{enumerate*}
Note that even though \sysname reduces the batch size, the overall system throughput or latency improves because of faster processing of requests in a batch due to prefix caching or speculative decoding.

Additionally, we observe that convergence times for the autoscaler are less than a few minutes as shown in Figure~\ref{fig:convergence_time}.
For example, Llama 70B with interactive SLO configuration has the maximum convergence time of $\sim$ 150 seconds.
As convergence times are low compared to the average instance duration, they are amortized as the cluster continues to operate.
Additionally, we note that smaller models have faster convergence times as the step time (i.e. time to observe local backpressure and adapt), is much lower compared to the larger models.
Consequently, Llama 8B has a convergence time of 15 seconds, which is 10$\times$ lower compared to Llama 70B.
Note that these experiments are done under constant arrival rate to achieve convergence and the convergence time is independent of the arrival rate.

\noindent
\textbf{Varying SLO Values.}
To understand the robustness of \sysname to varying SLOs, we set different values of ITL and TTFT SLOs in $W_A$ and $W_B$.
Such variation leads us to the following observations:
\begin{enumerate*}[label=(\alph*)]
    \item Increasing batch TTFT SLO leads to larger queue sizes and longer queue waiting time as shown in Figure~\ref{fig:queue_sizes}.
    As requests are queued longer, the opportunity for batch requests to multiplex with interactive requests via mixed instances increases leading to improved throughput.
    \item Large ITL SLOs lead to 100\% SLO satisfaction as shown in Table~\ref{table:itl_values}.
    Even with the maximum batch size, preemptions and increased attention computation costs do not violate ITL SLOs greater than 1 second.
    On the other hand, the probability of SLO violation with small ITL SLOs is less than 0.5\%.
    Such SLO violations can occur due to two reasons.
    First, the local autoscaler may exceed the ITL SLO target for a few steps if ITL measurements are noisy.
    Second, request evictions can cause the requests to be routed to another instance leading to repetition of the prefill step.
\end{enumerate*}

\noindent
\textbf{Queue Waiting Time Estimation Accuracy.}
Figure~\ref{fig:rwt_estimator} shows the coefficient of determination ($R^2$ values) when estimating waiting times with increasing queue sizes for the two models.
Overall, we confirm our observation that as the number of requests in the queue increase the waiting time estimation becomes more accurate.
Specifically, we find with 2000 requests, the estimation reaches an accuracy of 0.99.
While the estimation is highly accurate for longer request queues, it is not perfect.
When request queues are small, statistical averaging effects of continuous batching do not hold and the waiting time estimation is more conservative (i.e. higher) estimates leading to lower estimation accuracy.

\noindent \textbf{Impact of Arrival Burstiness.}
To understand the impact of arrival burstiness, we vary interactive request arrival rates in $W_A$ with a Gamma distribution as shown in Figure~\ref{fig:bursty_slos}.
The coefficient of variance (CV) represents burstiness of arrival rates. 
We set the default level of over-provisioning as 3 which is sufficient to handle burstiness of 8. 
However, as burstiness increases, request spikes become higher and the over-provisioning is insufficient to absorb bursts leading to SLO violations.
Hence, it is important to assign the over-provisioning level conservatively based on historical arrival patterns.

\noindent \textbf{Ablation Study.}
Figure~\ref{fig:ablation} shows the contribution of the local and global autoscaler to overall throughput improvements in \sysname.
To do so, we perform ablation studies where we replace
\begin{enumerate*}[label=(\alph*)]
    \item the global autoscaler with a utilization-based autoscaler (labeled as Global), and
    \item the local autoscaler with static batch sizes(labeled as Local).
\end{enumerate*}
Overall, we find that each autoscaler individually contributes between 30--60\% improvement in throughput for both interactive and batch requests.

%% file: 0008-related.tex
\section{Related Work}
\label{s:related_work}


\noindent
\textbf{General ML Model-Serving Systems.}
Traditional model-serving systems provide functionalities such as scheduling, placement, batching, and autoscaling.
Clipper~\cite{crankshaw2017clipper}, TensorFlow-Serving~\cite{tensorflow-serving}, MArk~\cite{zhang2019mark}, InferLine~\cite{inferline}, SHEPHERD~\cite{zhang2023shepherd}, and Clockwork~\cite{gujarati2020serving} are some earlier work on serving traditional ML models like ResNet that are relatively small.
INFaaS~\cite{romero2021infaas} and Cocktail~\cite{gunasekaran2022cocktail} propose a model-less serving framework to automate the model selection and autoscaling to meet SLOs.
However, they fail to consider the autoregressive property of LLMs leading to suboptimal throughput.

\noindent
\textbf{LLM Scheduling Optimization.}
Existing state-of-the-art LLM serving systems~\cite{yu2022orca,kwon2023efficient,tgi,triton} adopts continuous batching and a first-come-first-serve (FCFS) scheduling policy that suffers from head-of-line (HOL) blocking, which we address in \sysname.
FastServe~\cite{wu2023fast} proposes preemptive scheduling with a Multi-Level Feedback Queue. 
Andes~\cite{liu2024andes} defines Quality-of-Experience (QoE) for LLM serving as token delivery speed, and proposes a preemptive scheduler that maximizes QoE.
\sysname is the first autoscaling framework that optimizes SLO attainment while improving LLM-serving throughput and device utilization.

\noindent
\textbf{LLM Serving Backend Optimization.}
Various LLM serving backend optimization techniques have been proposed to improve token generation throughput and memory cost while adapting to fine-tuning paradigms such StreamingLLM, Speculative Decoding, ChunkedAttention, FlashAttention and more~\cite{liu2023deja,sheng2023flexgen,llm-inference,zhang2023h2o,liu2023scissorhands,ge2024model,miao2023spotserve,abhyankar2024apiserve,zhong2024distserve,patel2023splitwise,sheng2023slora,zhu2024relayattention,fang2021turbotransformers,leviathan2023fast}.
These backend LLM-serving optimizations are complementary to \sysname.

%% file: 0009-conclusion.tex
\section{Conclusion}
\label{s:conclusion}

We presented \sysname, a hierarchical autoscaling solution for LLM serving.
Evaluation using real-world LLM serving datasets on GPU devices demonstrate that \sysname improves SLO attainment by up to 90\%,  serving throughput up to 300\%, and reduces resource requirement up to 70\%.

%% file: appendix.tex
\appendix
\section{Appendix}

\subsection{Impact of Model Loading Times}
\label{s:discussion}


When model loading times are greater than the TTFT SLO for interactive requests, over-provisioning is a must to meet SLOs.
Given such over-provisioning exists, \sysname's global autoscaler can improve overall utilization by multiplexing the additional capacity with batch requests.

If model loading times are below the TTFT SLO (such as for models with less than 3B parameters), then \sysname can follow a completely elastic autoscaling approach similar to traditional cloud autoscaling.
In such a case, the need for over-provisioning and the global autoscaler reduces.
However, batch size adaptation with the local autoscaler continues to remain useful.

\subsection{Example Autoscaling Workflow}
\label{s:examples}

\begin{figure}[t]
    \centering
    \includegraphics{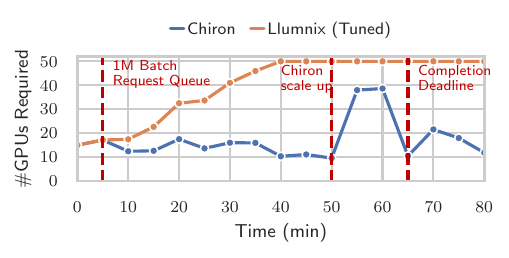}
    \caption{GPUs required over time for \sysname and Llumnix autoscalers when serving a batch and interactive workload for Llama 8B.}
    \label{fig:gpus_provisioned}
\end{figure}

\noindent \textbf{Initial Setup.}
In Figure~\ref{fig:gpus_provisioned}, we show an example workflow for \sysname and compare it against Llumnix.
Initially, at t=0 mins, the workload comprises only of interactive requests arriving with a Gamma distribution with mean of 30 requests/s and CV of 4.
Both \sysname and Llumnix would be over-provisioned in this scenario with an average of 15 GPUs.
Note that we use the tuned version of Llumnix which has similar instance-level throughput as \sysname.

\noindent \textbf{Arrival of Batch Requests.}
At t=5 mins, the batch request queue is populated with 1 million requests.
Llumnix does not enable queuing for these batch requests and immediately starts adding instances over time to reduce GPU utilization until the maximum cluster capacity of 50 instances is reached.
On the other hand, \sysname would maintain batch requests in the queue and prefer to multiplex with the over-provisioned capacity of 10 GPUs (out of 15 GPUs).
As batch requests have a relaxed ITL SLO, \sysname's local autoscaler is able to maintain a higher throughput of 20 requests/s on this over-provisioned capacity.

\noindent \textbf{Close to SLO Deadline.}
At t=50 mins, \sysname's waiting time estimation calculates that $\sim$ 200k requests still remain to be processed and 10 new instances are added to finish the queue by the deadline.
At t=65 mins, all requests are completed by \sysname.
As Llumnix does not adapt the batch size for the newly added instances, it continues to serve the requests at a reduced throughput.
Consequently, by the deadline of 65 mins, only 50\% of requests satisfy SLOs in case of Llumnix.

Overall, in this scenario, \sysname is able to use 60\% less GPU node hours while meeting all SLOs.